\documentstyle[preprint,epsfig,aps,tighten]{revtex}
\begin{document}
\def\sslash#1{#1\!\!\!/}
\def\slash#1{#1\!\!\!\!/}
\baselineskip 20pt
\noindent
\hspace*{13cm}
hep-ph/9812531\\
\noindent
\hspace*{13cm}
KIAS-P98051\\
\noindent
\hspace*{13cm}
SNUTP 98-149\\
\hspace*{13cm}

\vspace{1.8cm}

\begin{center}
{\Large \bf Bounds and Implications of Neutrino Magnetic Moments
from Atmospheric Neutrino Data}\\

\vspace*{1cm}

{\bf Sin Kyu Kang,$^{a,}$\footnote{skkang@ns.kias.re.kr}}
{\bf Jihn E. Kim,$^{a,b,}$\footnote{jekim@phyp.snu.ac.kr}} and
{\bf Jae Sik Lee$^{a,}$\footnote{jslee@ns.kias.re.kr}}\\

\vspace*{0.5cm}
 $^{(a)}$ School of Physics, Korea Institute for Advanced Study,
Seoul 130-012, Korea, and \\
 $^{(b)}$ Department of Physics and Center for Theoretical Physics, 
Seoul National University,\\ 
Seoul 151-742, Korea \\

(\today)
\vspace{0.5cm}

\end{center}

\def\hbar{{\mathchar'26\mskip-9muh}}

\begin{abstract}
\vspace{0.2cm}

\noindent
The neutral current effects of the future high statistics atmospheric 
neutrino data can be used to distinguish the mechanisms between  
a $\nu_\mu$ oscillation to a tau neutrino or to a sterile neutrino.
However, if neutrinos possess large diagonal and/or transition 
magnetic moments, the neutrino magnetic moments can contribute
to the neutral current effects which can be studied by the
single $\pi^0$ production events in the Super-K data. 
This effect should be included
in the future analyses of atmospheric data in the determination of
$\nu_\mu$ to tau or sterile neutrino oscillation.
\end{abstract}

PACS number(s): 13.15.+g, 14.60.Pq, 14.60.St 

\newpage
\section{Introduction}

Neutrinos might possess two properties which are feeble, but
will be important barometers of physics beyond the standard model
scale. These are neutrino mixings (masses and oscillations) \cite{pont} and
neutrino magnetic moments \cite{kim}. Before 1998, experiments gave
bounds on these properties, in general.  

But the recent results from the Super-Kamiokande 
Collaboration \cite{superk1} have 
provided a strong evidence for a deficit in the flux of atmospheric neutrinos, 
which are presented in the form of the double ratio
\begin{equation}
R=\frac{(N_{\mu}/N_e)_{obs}}{(N_{\mu}/N_e)_{MC}},
\end{equation}
which implies the existence of $\nu_\mu$ oscillation.
The measured value of $R$ for Super-Kamiokande is $0.61\pm 0.06 \pm 0.05$ for
the sub-GeV data and $0.67\pm 0.06 \pm 0.08$ for the multi-GeV data,
while we expect $R=1$ in a world without oscillations.
Muon neutrino oscillation into another species of neutrino provides a natural
explanation for the deficit and even the zenith angle dependence.
 The $\nu_{\mu} \rightarrow \nu_{\tau}$ oscillation is the most favorable
solution for the atmospheric neutrino problem, whereas the $\nu_{\mu} 
\rightarrow \nu_{e}$ oscillation is strongly disfavored by CHOOZ results 
\cite{chooz}. 
The oscillations into sterile neutrinos ($\nu_s$) 
give a plausible solution as well \cite{cjs}.
This evidence for the neutrino oscillations is also supported by the
SOUNDAN2 \cite{soudan} and by the Super-Kamiokande \cite{superk2} and MACRO
\cite{macro} data on upward-going muons.

It is usually assumed that the
neutral current effect in neutrino oscillation experiment
is unchanged, since any standard model neutrino produced
by oscillation has the same neutral current (NC) interaction. 
Therefore, the ratio of the neutral (NC) and charged currents (CC) events
is important to investigate the neutrino neutral current. 
Thus the observation of single $\pi^0$ events, induced by the 
neutral current, by Super-Kamiokande Collaboration 
\cite{superk1} can lead to important physical implications.

Since the $\pi^0$ NC event is detected as two 
diffuse rings whereas the  CC events due
to $\nu_e$ are detected as one diffuse ring due to $e^{\pm}$ and one sharp 
ring due to $\pi^{\pm}$ and the CC events due to $\nu_{\mu}$ are detected as 
two sharp rings from $\mu^{\pm}$ and $\pi^{\pm}$ \cite{pion1}.
Thus, a NC event can be discriminated from a $\nu_e$ CC event and a 
$\nu_{\mu}$ CC event \cite{smirnov1,sankar}. 
It has been considered difficult to separate NC and CC events clearly. But 
the single $\pi^0$ events described above can be used to discriminate
the NC events from CC events. Indeed, it is believed that the 
cleanest way to identify NC events in Super-Kamiokande is to detect a 
single $\pi^0$ from the process   $\nu + N \rightarrow \nu + N + \pi^0$, 
with $N$ being either a neutron  or a proton below Cerenkov threshold.
The $\pi^0$ is detected via its decay into two photons which lead to two 
diffuse $e$-like rings whose invariant mass is consistent with 
the $\pi^0$ mass \cite{pion1}.
The ratio
of $\pi^0$-like events to $e$-like events compared to the same ratio of the
Monte Carlo in the absence of the oscillation has been measured by
the Super-Kamiokande Collaboration \cite{superk2},
\begin{equation}
R_{\pi^0/e}=\frac{(\pi^0/e)_{data}}{(\pi^0/e)_{MC}} = 
            0.93\pm 0.07_{stat}\pm 0.19_{sys},
\end{equation}
where the systematic error is dominated by the poorly known single 
$\pi^0$ cross section, and the statistical error is based on 535 days 
of running.
The ratio $R_{\pi^0/e}$ is expected to be 1 for $\nu_{\mu} - \nu_{\tau}$ 
oscillations while 0.75 for $\nu_{\mu} - \nu_{s}$ oscillations  or
$\nu_{\mu} - \nu_{e}$ oscillations if one takes the measured 
$\nu_{\mu}/\nu_{e}$ ratio to be 0.65 \cite{smirnov1}.
The admixture of $\nu_{\mu} - \nu_{\tau}$ and $\nu_{\mu} - \nu_e$ 
oscillations also leads to a deviation of $R_{\pi^0/e}$ from 1.
Therefore, a precise measurement of the ratio will be used to
distinguish $\nu_\mu\rightarrow \nu_\tau$ from $\nu_\mu\rightarrow \nu_s$
oscillation.

At first sight, it is likely that any deviation of 
$R_{\pi^0/e}$ from 1 implies muon neutrino oscillation into 
a sterile neutrino. However, if there exists a large
muon neutrino magnetic moment (diagonal or transition), it will
produce an additional neutral current effect which has to be 
separated out to draw a definite conclusion. Indeed, right after
the discovery of the neutral current, the upper bound on
the muon neutrino magnetic moment was given \cite{ncnmm}.
Also, the experimental bounds on transition magnetic moments
and other properties in view of NC data were presented \cite{trans}.

The theoretical problem of obtaining a large neutrino magnetic
moment has begun with interactions beyond the standard model
\cite{kim}. In general, the loop diagram will have a (mass)$^2$ 
suppression, presumably by $M_X^{2}$ where $M_X$ can be the
$W$ boson mass or a scalar mass. It is possible to have a
large Dirac neutrino magnetic moment if the loop contains a
heavy fermion \cite{kim},
\begin{equation}
\mu_\nu\sim \frac{m}{M^2_X}
\end{equation} 
where $m$ is the mass of the heavy fermion.
This mechanism can be generalized in models with scalars 
\cite{scalar}. 

But, the same loop without the external
photon line would give a contribution to the neutrino mass
matrix. Therefore, one expects, taking the coupling
as $10^{-3}$, 
\begin{equation}
\mu_\nu\sim 10^{-3}\frac{m_e}{M_X}\left(\frac{m_\mu}{M_X}\right)^{1/3}\mu_B
\sim 10^{-13}\mu_B
\end{equation} 
where $\mu_B=e\hbar/2m_ec$ is the electron Bohr 
magneton and we 
used $m_\nu\sim $ O(1) eV for the numerical 
illustration.   To suppress the contribution to
the mass and still allow a large magnetic moment, continuous 
\cite{leurer} and discrete \cite{babu} symmetries have been
considered. In this case, the neutrino magnetic moment can be
as large as $\mu_{\nu_\mu}\sim 10^{-10}\mu_B$, which is not affected
by the SN1987A constraint $\sim 10^{-13}\mu_B$ \cite{goyal} 
since this bound applies to the electron neutrino only. 

However, a large transition magnetic moment to a sterile neutrino is not
forbidden that severely. For example, one can introduce
a transition moment with an accompanying mass as large
as several hundred MeV. Of course, the masses of the light neutrinos
are bounded by eV. In this case, the transition neutrino magnetic
moments can be as large as $10^{-7}\mu_B$ and may contribute
to NC events.
In particular, we are interested in
the single $\pi^0$ production through a large transition magnetic
moment,  which would contribute to $R_{\pi^0/e}$.
In this spirit, we will obtain the lower bound on 
the transition neutrino magnetic moment (to a sterile neutrino)
from $R_{\pi^0/e}$.
 
This paper is organized as follows:
In Sec. 2, we describe the amplitude for the single $\pi^0$ production.
In Sec. 3, the kinematics and the differential 
cross section for the single $\pi^0$ production are given.
In section 4, we present the contribution to the cross section of the
$\pi^0$ production generated by a possible neutrino
transition magnetic moment. In Sec. 5, we discuss the
physical implications based on the numerical result.

\section{Production of the single neutral pion}

The single $\pi^0$ production has two contributions: one from the
production and decay of the (3/2,3/2) baryon resonances and the
other from the continuum contribution. At low energies ($E<2$~GeV),
the contribution from baryon resonance production is a dominant
one for a single $\pi^0$ production \cite{fogli}:
\begin{eqnarray}
 \nu + N &\rightarrow & \nu + N^{\ast}, \nonumber \\
 N^{\ast} & \rightarrow & \pi^0 + N, 
\end{eqnarray}
where $N^{\ast}$ represents baryon resonances.
The cross section for the single $\pi^0$ production in the region
$W<1.6$ GeV ($W$ is the hadronic invariant mass in the final state) 
can be described following Fogli and Nardulli \cite{fogli}.
The effective Lagrangian for the neutrino neutral current is
defined by
\begin{equation}
{\cal L}_{NC}= {\frac{1}{\sqrt{2}}} G_F \bar{\nu}\gamma^{\lambda} 
(1+\gamma_5)\nu J^{NC}_{\lambda},
\end{equation}
by assuming the following general $V,A$ structure of the hadronic part 
of the NC:
\begin{equation}
J^{NC}_{\lambda} = g^3_V V^3_{\lambda} + g^3_A A^3_{\lambda} 
                 + g^8_V V^8_{\lambda}
                 +g^8_A A^8_{\lambda} + g^0_V V^0_{\lambda} 
                 + g^0_A A^0_{\lambda},
\end{equation}
where
$V^{i}_{\lambda}, A^{i}_{\lambda} (i=3,8,0)$ are the SU(3) nonet partners of
the CC \cite{review}. 
Neglecting the strange and charm NC's, one can write
\begin{equation}
J^{NC}_{\lambda} = g_V V^3_{\lambda} + g_A A^3_{\lambda} + g^{\prime}_V 
            V^{\prime 0}_{\lambda}+g^{\prime}_A A^{\prime 0}_{\lambda}, 
\end{equation}
where
\begin{eqnarray}
V^{\prime 0}_{\lambda} &=& \sqrt{\frac{1}{3}} 
(V^8_{\lambda} + \sqrt{2}V^0_{\lambda}), \\
A^{\prime 0}_{\lambda} &=& \sqrt{\frac{1}{3}} 
(A^8_{\lambda} + \sqrt{2}A^0_{\lambda}),
\end{eqnarray} 
the electromagnetic current being given by
$J^{em}_{\lambda} = V^3_{\lambda} + \frac{1}{3}V^{\prime 0}_{\lambda}$.
{}From the Weinberg-Salam model \cite{review}, 
\begin{eqnarray}
g_V &=& \frac{1}{2}-\sin^2 \theta_W, ~~~g_A=\frac{1}{2}, \\
g^{\prime}_V &=& -\frac{1}{3}\sin^2 \theta_W, ~~~g_A^{\prime}=0.
\end{eqnarray}

The isospin decomposition of one $\pi^0$ channels is given by
\begin{eqnarray}
 A(\nu p \rightarrow \nu p \pi^0) &=& \frac{1}{3}(2A_3+A_1) 
         + \sqrt{\frac{1}{3}}S, \\
A(\nu n \rightarrow \nu n \pi^0) &=& \frac{1}{3}(2A_3+A_1) 
         - \sqrt{\frac{1}{3}}S.
\end{eqnarray}
The reduced matrix elements $A_1, A_3$ are given by
\begin{eqnarray}
A_3 &=&\frac{1}{\sqrt{2}}(A^0_{\Delta} + A^0_{\pi} + A^0_N), \\
A_1 &=& \frac{3}{2\sqrt{2}}A^0_{NN\pi} - \sqrt{2}A^0_{\pi} -\frac{1}{2\sqrt{2}}
A^0_N + A^0_S + A^0_P + A^0_D,
\end{eqnarray}
where $A^0_{i}, i=N, NN\pi, P, S, D$ are given in the appendix and
the indices $S, P,D$ denote $S_{11}, P_{11}, D_{11}$
\cite{book}, respectively.
The contributions to the amplitude $S$ come from the following amplitudes
\begin{equation}
S=\frac{1}{2}\sqrt{3} \left(\sqrt{\frac{1}{2}} A^0_N+
   \sqrt{\frac{1}{2}} A^0_{NN\pi}
+\frac{2}{3}A^0_{P}+\frac{2}{3}A^0_{S}+ \frac{2}{3}A^0_{D}\right).
\end{equation}

As is well known, the dominant contribution to the amplitude $A$ comes from  
the $\Delta$ resonance in this region ($W<1.6$ GeV) \cite{smirnov2}.
Then, the amplitude for the single $\pi^0$ production can be described by 
\begin{equation}
A_{NC} \simeq \frac{G_F}{\sqrt{2}}\l_{\alpha}J^{\alpha}
\end{equation}
where
\begin{eqnarray}
&l_{\alpha} = \bar{u}(k^{\prime})\gamma_{\alpha} (1+\gamma_5)u(k), \\
&J^{\alpha} =  \frac{g}{m_{\pi}}\bar{u}(p^{\prime})q^{\rho}_\pi
            D_{\mu \rho}
           \left[-(g^{\mu \alpha}\slash{Q}-Q^{\mu}
           \gamma^{\alpha})\gamma_5g_V\frac{C_3^V}{M_N}\right.\nonumber\\ 
         &~~~~~\ \ \ \ \ \ \ \left.- (g^{\mu \alpha}Q\cdot p^{\prime}
           -Q^{\mu}p^{\prime \alpha})\gamma_5 g_V
           \frac{C_4^V}{M_N^2}-g^{\mu \alpha}
           g_AC_5^A \right]u(p),
\end{eqnarray}
where $M_N$ is the nucleon mass, 
$D_{\mu \rho}$ is the propagator of Rarita-Schwinger field which is given by
\begin{equation}
D_{\mu \rho}(p)=\frac{\sslash{p}+M^{\prime}}{p^2-M^{\prime 2}+iM^{\prime}\Gamma}
\left(g_{\mu \rho}-\frac{2}{3}\frac{p_{\mu}p_{\rho}}
{M^{\prime 2}}
+\frac{1}{3}\frac{p_{\mu}\gamma_{\rho}-p_{\nu}\gamma_{\rho}}
{M^{\prime}} - \frac{1}{3}\gamma_{\mu} \gamma_{\rho}\right),
\end{equation}
and $q=p^{\prime}+q_\pi-p=k-k^{\prime}$ is the momentum transfer,
$\Gamma $ is the decay width,  $C_i^V$ and
$C_i^A (i=3,4,5)$ are the vector and axial vector transition form factors as
defined by Llewellyn-Smith \cite{smith}, and $M^\prime$ is mass of
the $\Delta$ resonance.
As shown in Ref.\cite{fogli}, the form factors $C_i(q^2)$ can be obtained by
comparison with the values of the helicity amplitudes given by the relativistic
quark model \cite{quark}. The explicit forms are given by
\begin{eqnarray}
C_3(q^2) &=& \frac{1.7\sqrt{1-q^2/4M_R^2}}{[1-q^2/(M_R+M_N)^2]^{3/2}
        [1-q^2/0.71\ {\rm GeV}^2]^2}, \\
C_4(q^2) &=& -\frac{M_N}{\sqrt{W^2}}C_3(q^2), \\
C_5(q^2) &=& 0.
\end{eqnarray}
Then, the vector form factors $C_i^V(q^2)$ used in Eq.(20) are given by
\begin{equation}
C_i^V(q^2)=\sqrt{3}C_i(q^2).
\end{equation}
The axial form factors $C^A_i(q^2)$ is also
taken to be the general formula
\begin{equation}
C^A(q^2)=\frac{C^A(0)}{(1-q^2/M^2_A)^2},
\end{equation}
where $M_A=0.65~\mbox{GeV}$.

\section{Kinematics}

Consider the process given in (5). It is convenient
to choose the center of momentum frame. Without loss of 
generality, we can choose the initial four momenta
of the neutrino and the nucleon in the CM frame as
$(p,p,0,0)$ and $(E_N,-p,0,0)$, respectively, where
\begin{eqnarray}
p &=& \frac{E_{\nu}}{\sqrt{1+2E_{\nu}/m_N}}, \\
E_N &=& \frac{m_N+E_{\nu}}{\sqrt{1+2E_{\nu}/m_N}},
\end{eqnarray}
where $E_{\nu}$ is the incident neutrino energy in the LAB
frame.
The final four momenta of neutrino, nucleon and $\pi^0$
are $k^{\prime},p^\prime$ and $q_\pi$, respectively,
where
\begin{eqnarray}
k^{\prime} &=& E_{\nu^{\prime}}\left[1, (\cos \theta,
 \sin \theta, 0)\right], \\
p^{\prime} &=& E_{N^{\prime}}\left[1, \sqrt{1-\frac{m^2_{N}}{E_{N^{\prime}}^2}}
        (-\cos \beta \cdot\cos \theta+\sin \beta \cdot \sin \theta \cdot
         \cos \phi, \right. \nonumber \\
& &\left. -\cos \beta \cdot\sin \theta -\sin \beta\cdot
        \cos \theta \cdot \cos \phi, -\sin \beta \cdot \sin \phi)
        \frac{}{}\right],\\
q_\pi &=& E_{\pi}\left[1, \sqrt{1-\frac{m^2_{\pi}}{E_{\pi}^2}}
        (-\cos \alpha \cdot\cos \theta-\sin \alpha \cdot \sin \theta \cdot
         \cos \phi,\right.\nonumber \\
& & \left.-\cos \alpha \cdot\sin \theta +\sin \alpha\cdot
        \cos \theta\cdot \cos \phi, \sin \alpha \cdot \sin \phi)\frac{}{}\right], 
\end{eqnarray}
where
\begin{eqnarray}
\cos \alpha &=& \frac{\vec{k}^{\prime 2}-\vec{p}^{\prime 2}+\vec{q}_\pi^2}
                {2|\vec{k}^{\prime}|\cdot |\vec{q}_\pi|}, \\
\cos \beta &=& \frac{\vec{k}^{\prime 2}+\vec{p}^{\prime 2}-\vec{q}_\pi}
                {2|\vec{k}^{\prime}|\cdot |\vec{p}^{\prime}|}.
\end{eqnarray}
The angles $\theta, \phi$ correspond to the rotations around
$z-$ and $x-$ axes, respectively.
Then, the differential cross section $d\sigma$ can be expressed as
\begin{eqnarray}
d\sigma &=& \frac{(2\pi)^4|\overline{A_{NC}}|^2}{4(p\cdot k)}
         d\Phi_3 \nonumber \\
        &=&\frac{|\overline{A_{NC}}|^2}{32 (2\pi)^4 M_N E_{\nu}}
       dE_{\nu^{\prime}} dE_{\pi}d\phi d(\cos \theta),
\end{eqnarray}
where
\begin{equation}
|\overline{A_{NC}}|^2=\frac{G_F^2}{2} L_{\mu \nu}J^{\mu \nu},
\end{equation}
with
\begin{equation}
L_{\mu \nu} = \frac{1}{2}\sum_{\mbox{spins}}l^{\dagger}_{\mu}
                     l_{\nu},
\end{equation}
and 
\begin{equation}
J_{\mu \nu} = \frac{1}{2}\sum_{\mbox{spins}}  J^{\dagger}_{\mu} J_{\nu}
\end{equation}
where the summation is performed over the hadronic spins.

\section{Contribution from the neutrino magnetic moment}

In this section, we consider the $\Delta$ production arising
from the Feynman diagram shown in Fig.~1. The decay of $\Delta$
to a nucleon plus $\pi^0$ follows in the detector, and one
observes the $\nu^\prime+N+\pi^0$ final state. The $\nu_\mu-
\nu^\prime-\gamma$ vertex is parametrized by a transition
magnetic moment 
\begin{equation}
if_{\nu^\prime\nu_\mu}\mu_B\bar u(l^\prime)_{\nu^\prime} 
\sigma_{\mu \nu} q^{\nu}u(l)_{\nu_\mu},
\end{equation}
where $q=l-l^{\prime}=p^{\prime}+q_\pi-p$ is the momentum transfer. 
The coupling $f_{\nu^\prime\nu_\mu}$ at $q^2=0$ is the
transition neutrino magnetic moment in units of the
electron Bohr magneton and will be denoted as $f^\prime$.

The squared matrix element that describes the 
single $\pi^0$ production, induced by
a transition neutrino magnetic moment $f^\prime$, can be written as
\begin{equation}
|\overline{A}_M|^2=\frac{f^{\prime 2}\mu_B^2}{q^4}M^{\mu \nu} 
                  J^{em}_{\mu \nu},
\end{equation}
where
\begin{equation}
 M^{\mu \nu} = \frac{1}{2}\sum_{\mbox{spins}}
 [\bar{u}(l^{\prime})\sigma^{\mu \alpha}q_{\alpha}u(l)]
 [\bar{u}(l)\sigma^{\nu \beta}q_{\beta}u(l^{\prime})]
\end{equation}
and
\begin{equation}
J^{em}_{\mu\nu}=\frac{1}{2}\sum_{\mbox{spins}}J^{em}_\mu J^{em}_\nu
\end{equation}
where $J^{em}_{\mu \nu}$ is the hadronic electromagnetic current 
given before by $J^{em}_\mu=V^3_\mu+(1/3)V^{\prime 0}_\mu$.

\section{Results and Discussions}

In this section we present the numerical results of the cross section
of the single $\pi^0$ production for the NC interactions.
The calculated cross section generated by the neutrino transition magnetic
moment is shown in Fig.~2 as a function of the incident neutrino energy
for the hadronic invariant mass less than $1.6$ GeV.
We find that the values of the cross 
section is of the order $10^{-40}~{\rm cm}^2$ for $E_{\nu}\leq 2$ GeV.

In order to see how the contribution to the cross 
section generated by the neutrino magnetic moment can be constrained by 
the experimental results of the ratio $R_{\pi^0/e}$, 
it is sufficient to calculate the ratio $\frac{\sigma_{f^\prime}}
{\sigma^{NC}_{0}}$ where $\sigma_{f^\prime}, \sigma^{NC}_{0}$
are the cross sections  of the single $\pi^0$ production 
from the neutrino magnetic moment and the standard model NC
interactions, respectively.
The reason is that $e$-like events (CC) is not affected by the presence of
the neutrino magnetic moment.
In Fig.~3, we plot this ratio
\begin{equation}
r_{f^\prime/NC}=\frac{\sigma_{f^\prime}}{\sigma^{NC}_0}
\end{equation}
as a function of the incident neutrino energy $E_\nu$ 
for $f^\prime=f_0^\prime\equiv 0.6\times 10^{-8}$. 
$f_0^\prime$ is defined as the value giving a 
similar contribution as the NC interaction. If $f^\prime$ is $\epsilon$
times $f_0^\prime$, Fig.~3 should be multiplied by a
factor $\epsilon^2$.  Note that the contributions from
the transition magnetic moment and from the standard model NC do not
mix in the process $\nu+N\rightarrow\nu^\prime+N^\prime$ 
due the unmixable $\gamma$ matrix structure among these two. 
However, for $\nu+N
\rightarrow\nu^\prime+\Delta$ there are terms which mix these
two contributions. This is because the form factors $\Gamma_\mu$, 
defined in $\bar\Delta\Gamma_\mu N$,
can match the $\gamma$ matrices through taking out one index
by $q_\mu$.  
The dashed line corresponds to the case of no cut whereas the solid line
corresponds to the hadronic invariant mass cut at $1.6$ GeV.
The current experimental result of the ratio $R_{\pi^0/e}$
implies that the possible excess from 1 amounts to 0.13 from which
we can obtain the constraint on the neutrino magnetic moment.
For example, at $E_{\nu}=5$ GeV a constraint
$r_{f^\prime/NC} \leq 0.13$
leads to
\begin{equation}
f^\prime \leq 2.2\times 10^{-9}.
\end{equation}

The transition magnetic moment of this magnitude implies the
muon neutrino and sterile neutrino mass matrix of the form
\begin{equation}
M_{\nu_\mu \nu^\prime}=\left(\matrix{m_{11},\ m_{12}\cr
                                     m_{21},\ m_{22}}\right)
\end{equation}
where $m_{12}\sim m_{21}$ is roughly $10^4$ times $m_{11}\sim 
O(10^{-2})$~eV. Thus $m_{12}$ is of order $\le$ 100~eV. The diagonalization 
process should not change the mass of the muon neutrino
drastically, i.e. $m_{22}\ge m_{12}^2/m_{11}\sim 1$~MeV. 
Therefore, a
singlet neutrino at the intermediate scale with possible
beyond the standard model interactions (scalar or gauge)
can lead to a sizable transition magnetic moment.

The effects of the muon neutrino transition magnetic moment
should be separated out toward a final determination of
the muon neutrino oscillation to the tau neutrino or to a sterile neutrino.
The most promising method is to study the energy distribution
of the final $\pi^0$, since
the kinematics for the magnetic moment is different from the
NC interactions where the former has a $1/q^2$ dependence in
the differential cross section while the latter has no $q^2$
dependence at low energy. 

Indeed, if the transition magnetic moment is discovered
by the measurement of the energy distribution, it will
hint an intermediate scale physics. On the other hand,
one can compare this
anticipation to the earlier expectation that neutrinos must
oscillate due to the belief that singlet fermions, the
remnants of grand unification or the standard model superstring, would
be present at the intermediate scale \cite{seesaw,string}.
Similarly, if singlet neutrinos are present much above the eV scale,
there may be a large transition magnetic moment which can
be detected by the future high statistics atmospheric
neutrino experiments. 

\vspace{1cm}
\centerline{{\bf Acknowledgments}}
We thank S. Y. Choi for helpful discussions.
One of us (JEK) is supported in part by KOSEF, MOE through
BSRI 98-2468, and Korea Research Foundation. 

\newpage
\section*{Appendix}
We present the expressions of $A^0_i, ~~i=N,NN\pi, P, S, D$ \cite{fogli}:
\begin{eqnarray}
A^0_{\pi} &=&3\sqrt{\frac{1}{2}} G_F J^{\lambda}\sqrt{2}g_{NN\pi}\bar{u}
     (p^{\prime}) \gamma_5 u(p)\frac{2q^{\prime\lambda} + Q^{\lambda}}
     {(q^{\prime}+Q)^2-m^2_{\pi}}g^{\prime}_AF_{\pi}, \\
A^0_N &=& -3\sqrt{\frac{1}{2}}G_F 
      J^{\lambda}\sqrt{2}g_{NN\pi}\bar{u}(p^{\prime}) \nonumber \\
 &\times &\left[g^{\prime}_VF_1\gamma_{\lambda} + g^{\prime}_V
  \frac{F_2}{2M_N}[~\gamma_{\lambda},\slash{Q}~]-g^{\prime}_A\frac{1}{3}F_A
  \gamma_{\lambda}\gamma_5\right]\frac{\sslash{p}^{\prime}+\slash{Q}+M_N}{(p^{\prime}+Q)^2
  -M_N^2}
  \gamma_5u(p), \\
A^0_{NN\pi} &=& 3\sqrt{\frac{1}{2}}G_F J^{\lambda}\sqrt{2}g_{NN\pi}\bar{u}
      (p^{\prime})\frac{\sslash{k}+M_N}{k^2-M_N^2} \nonumber \\
    &\times &\left[g^{\prime}_VF_1\gamma_{\lambda} + g^{\prime}_V
    \frac{F_2}{2M_N}[~\gamma_{\lambda},\slash{Q}~]-g^{\prime}_A\frac{1}{3}F_A
    \gamma_{\lambda}\gamma_5\right]u(p), \\
A^0_{P} &=& -\frac{3}{2}\sqrt{\frac{1}{2}}G_F J^{\lambda}f_P\bar{u}
      (p^{\prime})\frac{\sslash{k}+M_P}{k^2-M_P^2}D^S_Ag^{\prime}_A
      \gamma_{\lambda}\gamma_5 u(p), \\
A^0_{S} &=& \frac{9}{2}\sqrt{\frac{1}{2}}G_F J^{\lambda}f_S\bar{u}
      (p^{\prime})\frac{\sslash{k}+M_S}{k^2-M_S^2} \nonumber \\
      &\times &\left[g^{\prime}_VG^S_1\gamma_{\lambda} - g^{\prime}_V
      \frac{G^S_2}{2M_N}[~\gamma_{\lambda},\slash{Q}~]\gamma_5
      +g^{\prime}_A\frac{1}{3}G^S_A
      \gamma_{\lambda}\right]u(p), \\
A^0_{D} &=& -\frac{9}{2}\sqrt{\frac{1}{2}}G_F \frac{f_D}{m_{\pi}}
        \bar{u}(p^{\prime}) q^{\prime \mu}\gamma_5 D_{\mu \rho}
      \left[(\slash{Q}~g^{\rho \lambda}-Q^{\rho}\gamma^{\lambda})g^{\prime}_V
      \frac{H^S_3}{M_N}+(k\cdot Q~g^{\rho \lambda}
      -Q^{\rho}k^{\lambda})g^{\prime}_V
      \frac{H^S_4}{M_N^2} \nonumber \right. \\
      &-& \left.(p\cdot Q~g^{\rho \lambda}-Q^{\rho}p^{\lambda})g^{\prime}_V
      \frac{H^S_5}{M_N^2} -g^{\rho \lambda}g^{\prime}_A
     \frac{1}{3}H^S_A\gamma_5\right]u(p) \nonumber \\
\end{eqnarray}
where $G_F=1.023\cdot 10^{-5}M_N^{-2}$ is the Fermi constant, 
$g_{NN\pi}/4\pi=14.8$ is the $NN\pi$ coupling constant, 
$M_N$ is the nucleon mass, $m_{\pi}$ is the 
pion mass and $M_R$ the mass of the generic resonance $R$.
We also assume $M_A=0.65~\mbox{GeV}/c^2 ~~(M_A=1.0~\mbox{GeV}/c^2$ for
the axial mass of the $P_{33}(P_{11}, S_{11}, D_{13})$ .
The generic vector form factor in the amplitudes Eqs.(45)-(47) is given by
the general formula
\begin{equation}
V(t) = V^{p}(t) + V^{n}(t),
\end{equation}
where $V^{p}(V^{n})$ is the electromagnetic form factor with a proton(neutron)
as target.

With regard to the pion and nucleon form factors, we use the following usual
forms,
\begin{eqnarray}
F_{\pi}(t) &=& \frac{1}{1-t/0.47\ {\rm GeV}^2}, \\
F_{1}^V(t) &=& \left(1-\frac{3.7t}{4M_N^2-t}\right)\frac{1}{(1-t/0.71
\ {\rm GeV}^2)^2}, \\
F_{2}^V(t) &=& 1.855\left(1-\frac{t}{4M_N^2}\right)^{-1}\frac{1}{(1
-t/0.71\ {\rm GeV}^2)^2}, \\
F_{A}(t) &=& 1.23\frac{1}{(1-t/0.81\ {\rm GeV}^2)^2}, 
\end{eqnarray}
where $t=Q^2$ is the moment transfer and
the axial form factor is characterized by $F_A(0)=1.23$, which is derived
from neutron decay, whereas $M_A=0.90~\mbox{GeV}/c^2$ 
is perfectly compatible with an overall fit to neutrino experiments.
The form factors $G_i$ and $H_i$ are explicitly given in Ref.\cite{fogli}.

The axial form factors in the amplitudes Eqs.(45)-(47)
are taken to be the general formula
\begin{equation}
A(t)=\frac{A(0)}{(1-t/M^2_A)^2}
\end{equation}
where $M_A=0.65~\mbox{GeV}$ for $C^A_5$ and $M_A=1.0~\mbox{GeV}$
for the other resonances.

\begin{figure}[h]
\hspace{1pt}
\vspace{75pt}
\begin{center}
\epsfig{figure=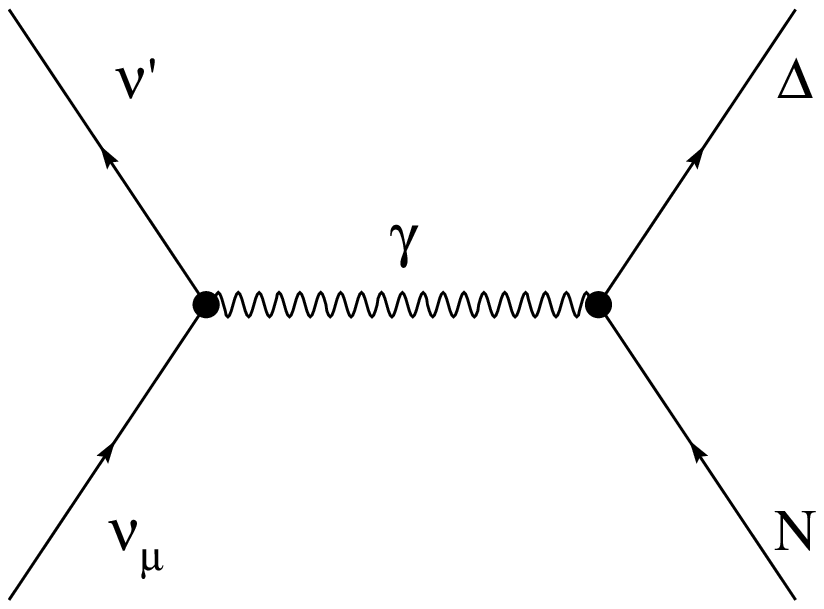}
\end{center}
\caption{Feyman diagram for the $\Delta$ production arising from the 
neutrino transition magnetic moment.}
\label{fig1}
\end{figure}

\begin{figure}[ht]
\hspace*{-1.0 truein}
\psfig{figure=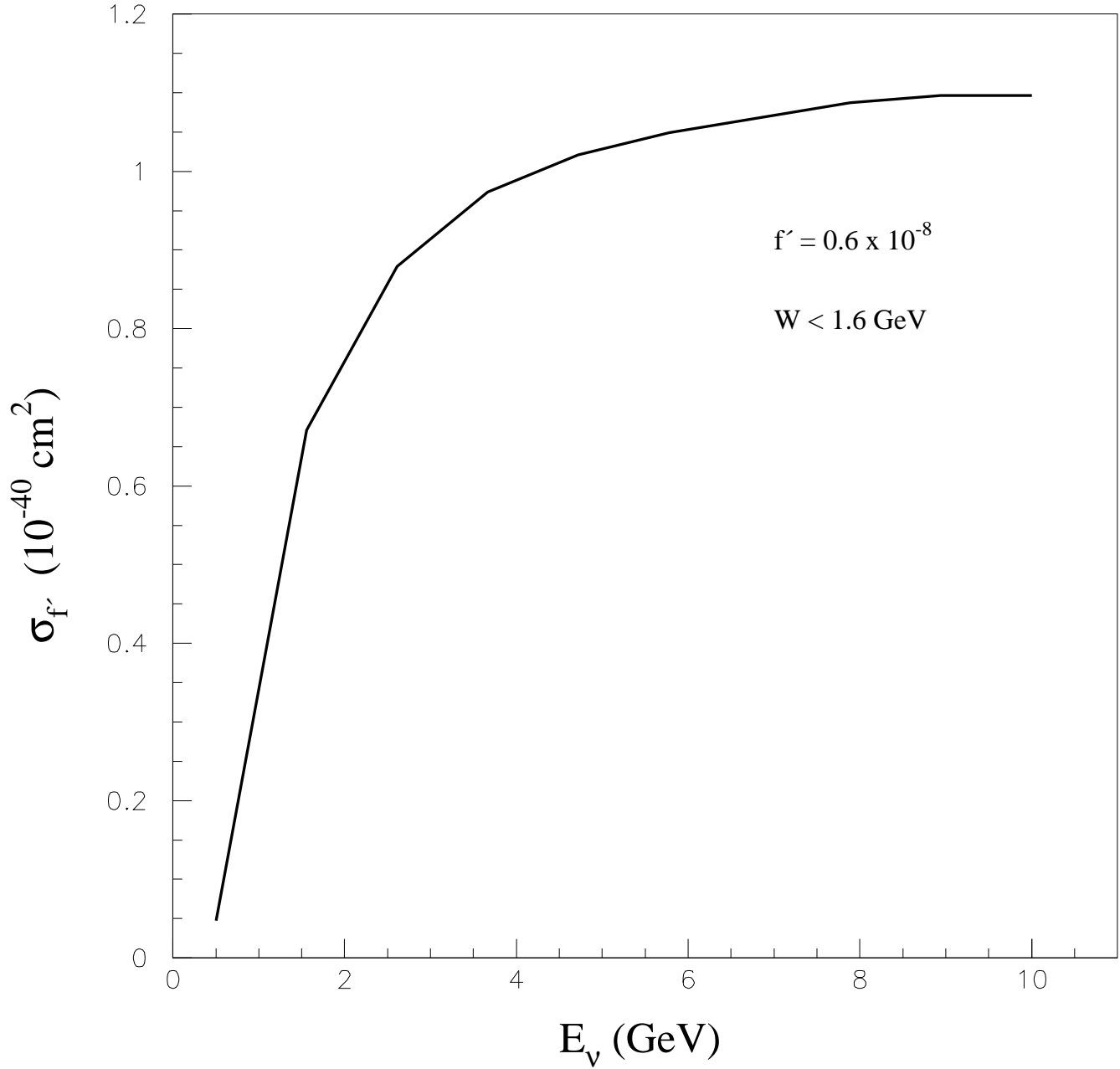}
\caption{
The cross section generated by the neutrino transition magnetic
moment as a function of the incident neutrino energy
for the hadronic invariant mass less than $1.6$ GeV.
$f^{\prime}$ is taken to be $0.6\times 10^{-8}$.}
\label{fig2}
\end{figure}

\begin{figure}[ht]
\hspace*{-1.0 truein}
\psfig{figure=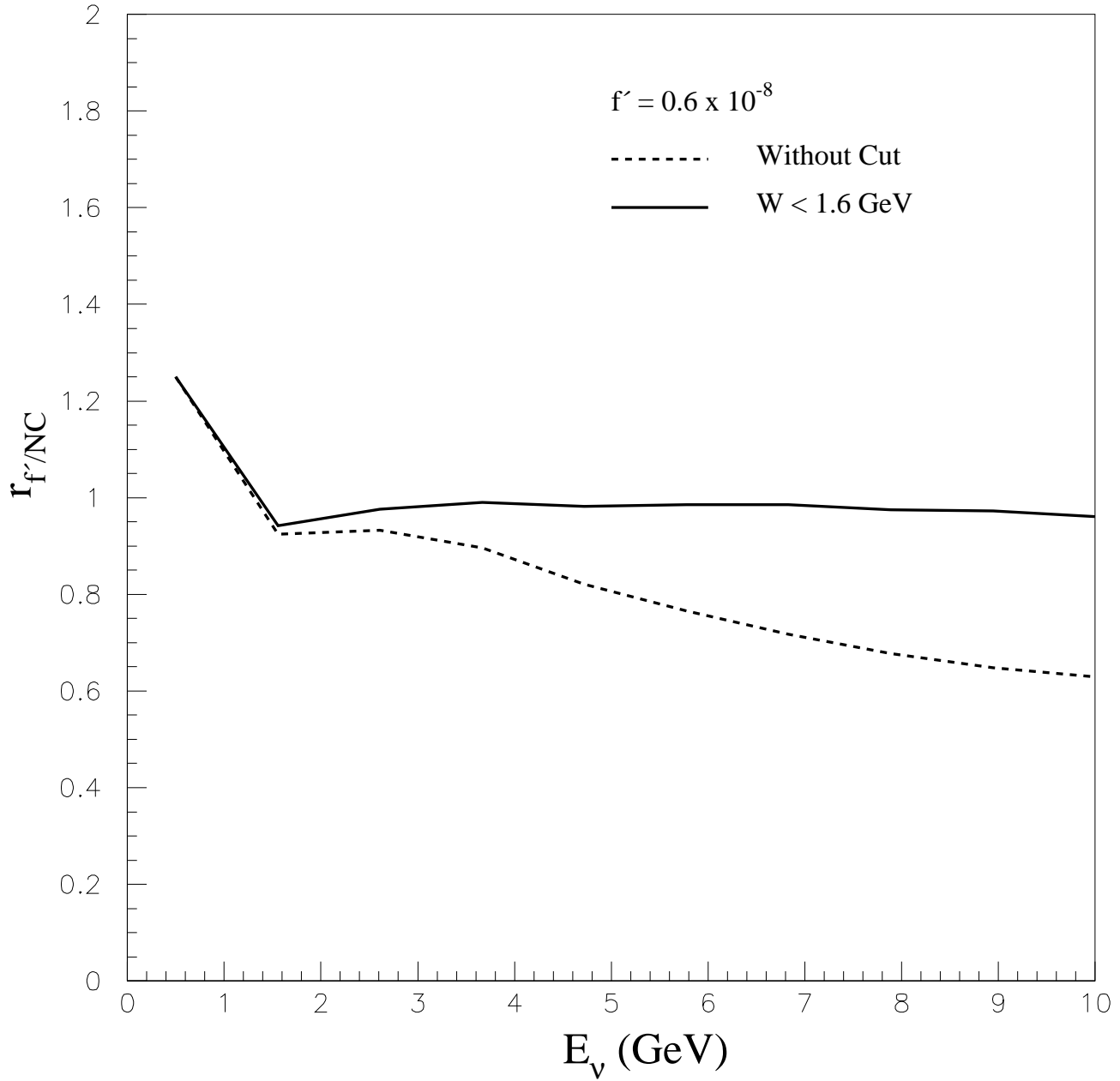}
\caption{
Plots of $r_{f^{\prime}/NC}$
as a function of the incident neutrino energy
for $f^{\prime}= 0.6\times 10^{-8}$.
The dashed line corresponds to the case of no cut and the solid line
corresponds to the invariant mass cut at 1.6 GeV}
\label{fig3}
\end{figure}

\end{document}